# Elastic beam algorithm for generating circular cartograms

Zhiwei Wei[a,b*], Su Ding[c], Wenjia Xu[d], Lu Cheng[e], Song Zhang[a,b], Yang Wang[a,b*]

[a] *The Aerospace Information Research Institute, Chinese Academic of Sciences, Beijing 100190, China*; [b]*Key Laboratory of Network Information System Technology (NIST), Aerospace Information Research Institute, Chinese Academy of Sciences, Beijing 100190, China*; [c]*College of Environmental and Resource Science, Zhejiang A & F University, Zhejiang 311300, China*; [d]*Beijing University of Posts and Telecommunications*; [e]*School of Resource and Environment Science, Wuhan University, Wuhan 430079, China.*

Address for correspondence: Zhiwei Wei. E-mail: 2011301130108@whu.edu.cn

Yang Wang. E-mail: Primular@163.com

# Elastic beam algorithm for generating circular cartograms

**Abstract**: The circular cartogram, also known as the Dorling map, is a widely used tool for visualizing statistical data. It represents regions as circles with their areas in proportion to the statistical values and requires circle displacement to avoid overlap and maintain spatial relationships. In this paper, we propose a new approach for circular cartogram production that utilizes the elastic beam displacement algorithm in cartographic generalization. First, the initial circles are generated with their areas in proportion to the statistical values. Second, an elastic beam structure is built as a proximity graph based on the spatial relations between the circles. Third, the circles violating the quality requirements are considered to have a force on the nodes of a beam. Fourth, the elastic beam algorithm is applied to assign forces for each node to determine the new positions of the circles. Steps two through four are repeated until a circular cartogram that meets the defined quality requirements is obtained. The experiments indicate that the proposed approach can successfully generate circular cartograms without overlaps while maintaining topology and relative relationships with higher quality than existing approaches.

**Keywords**: Cartogram; Dorling Map; Displacement; Population mapping; Energy minimization.

## 1. Introduction

Area cartograms are highly effective for visualizing geographic data through the proportional reshaping of regions based on their statistical values (Tobler, 2004; Nusrat and Kobourov, 2016). By utilizing area and shape as key graphic variables, area cartograms provide an intuitive way for map readers to comprehend spatial data quantity (Inoue, 2011).

A range of approaches has been proposed for automating the production of area cartograms, which can be broadly classified into complex shape-based or regular shape-based. In complex shape-based approaches, complex shapes are used to represent the geographic regions, preserving the topology relations between regions, which enhances understanding of spatial relations (Chrisman, 1977; Dougenik et al., 1985; Jackel, 1997; Du and Liu, 1999; Chrisman, 2002; Gastner and Newman, 2004; Wolf, 2005; Sun, 2013a; Sun, 2013b; Gastner et al., 2018; Kronenfeld, 2018; Sun, 2020). However, significant distortion can occur, making it difficult for users to distinguish area differences (Inoue, 2011). In contrast, regular shape-based cartograms use familiar shapes such as circles or rectangles to reshape geographic regions, which allows users more easily perceive the area differences (Raisz, 1934; Dorling, 1996; Heilmann et al. 2004; Speckmann et al, 2006; van Kreveld and Speckmann, 2007; Inoue, 2011; Buchin et al., 2012; Tang, 2013). If the spatial relations between regions are also well-considered, the spatial data can be better visualized with regular shape-based cartograms in this regard.

Circles are a common shape used in area cartograms that are known as circular cartograms or Dorling map after its creator Daniel Dorling (1996). Nowadays, the circular cartogram is widely adopted by data visualization libraries, such as Protovis (2010) and D3 (Bostock et al., 2011), which facilitate its implementation in web applications. The generation of a circular cartogram involves two main steps. First, the initial circles are generated with their areas proportional to the statistical values, and placed in the centers of their respective regions; second, the circles are displaced to avoid overlaps and maintain the spatial relations between regions. The circle displacement is the critical process in circular cartogram production, which has been addressed by different approaches. For instance, Dorling (1996) and Tang (2013) proposed a simulation-based approach, inspired by the orbits of celestial bodies, while Inoue (2011) used the least-squares adjustment (LSA) for global optimization. However, these approaches have some limitations. The simulation-based approach may lead to local optimization and cannot well maintain the topology relations between regions. Using the LSA, it is hard to describe all quality requirements with unified linear functions, and the solutions to which do not always exist. Therefore, it is essential to explore new approaches for circle displacement with all quality requirements satisfied as much as possible.

Displacement is a fundamental operation in cartographic generalization, and many algorithms have been developed to address this task. Among these algorithms, the elastic beam algorithm is particularly powerful because it enables global optimization of map quality requirements through energy minimization (Bader et al., 2001; Liu et al., 2014). The circles in a cartogram can also be considered as map objects. The requirements for circular cartograms, such as overlap avoidance and topology relation maintenance, are analogous to the map quality requirements. Therefore, the elastic beam algorithm used in cartographic generalization can also be applied to displace circles in circular cartogram production. By achieving global optimization of the quality requirements, the elastic beam algorithm can produce a circular cartogram that satisfies all quality requirements to the greatest possible extent.

Motivated by these thoughts, we utilized the elastic beam algorithm to displace circles for producing circular cartograms. The initial circles were generated with their areas being proportional to the statistical value first. We then connected the circles to form a proximity graph that represented their spatial relations. In this graph, each node represented a circle, and each edge linked two neighboring circles and represented their spatial relationships such as being adjacent to each other. The proximity graph was treated as an elastic beam structure, with the quality requirements for the circular cartogram being converted into attractive or repulsive forces on each circle. These forces were assigned to each circle to determine their new locations iteratively using the elastic beam algorithm until an acceptable result was achieved. It should be noted that "force" here stands for a target displacement and not for a physical force in the sense of Newtonian mechanics.

## 2. Related works

### 2.1 *Quality requirements for circular cartogram*

To ensure the readability, aesthetics, and accuracy of the represented information, a circular cartogram must satisfy specific quality requirements. These requirements can be categorized as follows:

(1) Geometry

Shape: Regions must be represented by circles (Dorling, 1996);

Area: Circle areas must be proportional to the statistical values (Dorling, 1996);

(2) Relation

Overlap avoidance: Circles must not overlap (Dorling, 1996);

Topology relation maintenance: The circles need to be adjacent to each other as much as possible if their corresponding regions are adjacent and vice versa (Dorling, 1996; Inoue, 2011);

Relative relation maintenance: The relative relations between regions need to be maintained as much as possible, particularly their relative orientations (Inoue, 2011);

(3) Distribution

Total displacement distance minimization: The displacement distance of circles needs to be minimized (Liu et al., 2014).

It should be noted that these quality requirements may conflict with or enhance one another. For example, displacement to avoid overlaps may increase the total displacement distance. As a result, the quality requirements need not be fully satisfied but should be fulfilled as much as possible. Additionally, some requirements may have higher priority than others. For example, overlap avoidance may have a higher priority than topology relation maintenance since the overlaps must be avoided, while the topology relations only need to be maintained as much as possible.

**2.2 *Algorithms for circular cartograms***

Circular cartograms have a long history, with one of the earliest examples being the population map in The Rand McNally World Atlas in 1897 (Reyes Nunez, 2014). This map reshaped geographic regions according to their statistical values and was widely adopted in newspapers and magazines because of its ability to convey data quantity and spatial distribution. However, early circular cartograms were mostly produced by hand. Dorling (1996) developed the first computer program for automated circular cartogram production. Because the initial circles in a circular cartogram can be easily generated according to the statistical values, the crucial aspect of circular cartogram production is displacing these circles to meet defined quality requirements. To achieve this, Dorling (1996) simulated the orbits of stars and planets to displace the circles. But the circles in his approach are displaced one by one, which results in local optimization only, and the topology relations and relative relations between circles are not well maintained. Inoue (2011) then attempted to achieve global optimization of quality requirements by using the least-square adjustment method. But it is hard to describe all involved quality requirements as unified linear functions if more regions or quality requirements are considered. Furthermore, a satisfactory solution may not always exist. Consequently, many modern data visualization libraries such as D3 (2013) and Protovis (2010) still employ Dorling's (1996) approach. To improve efficiency, Tang (2013) applied parallel computing by dividing the task into subproblems.

In summary, while Dorling's (1996) approach is still widely used for circular cartogram production, it falls short of meeting all quality requirements. Though some approaches using global optimization have also been introduced, they may not always be feasible. Hence, it is necessary to develop a new approach that can well balance all the quality requirements of circular cartograms.

**2.3 *The elastic beam algorithm***

The elastic beam algorithm was originally proposed for building displacement by Bader (2001) based on the principle of energy minimization and was later developed by Liu et al. (2014). The algorithm models the proximity graph of buildings as an elastic beam structure, where the proximity graph is constructed according to the spatial relations between buildings. Buildings that violate the map quality requirements are assigned a force that displaces them. Because the edges in the proximity graph are treated as elastic beams, the forces on the graph will cause the beams to stretch and bend. The forces and moments acting on each beam's two end nodes are represented as ($f_1^x$, $f_1^y$, $M_1$) and ($f_2^x$, $f_2^y$, $M_2$) (Figure 1). The end nodes are then displaced or rotated under these forces and moments, their displaced distances and rotated angles are represented as ($d_1^x, d_1^y, \theta_1$) and ($d_2^x, d_2^y, \theta_2$) (Figure 1). The moments ($M_1$ and $M_2$) on the two end nodes are defined as Eq. (1), where ($x_1$, $y_1$), and ($x_2$, $y_2$) are the coordinates of two end nodes (Liu et al., 2014).

$$\begin{cases} M_1 = (y_1 - y_2) * f_1^x + (x_1 - x_2) * f_1^y \\ M_2 = (y_2 - y_1) * f_2^x + (x_2 - x_1) * f_2^y \end{cases} \quad (1)$$

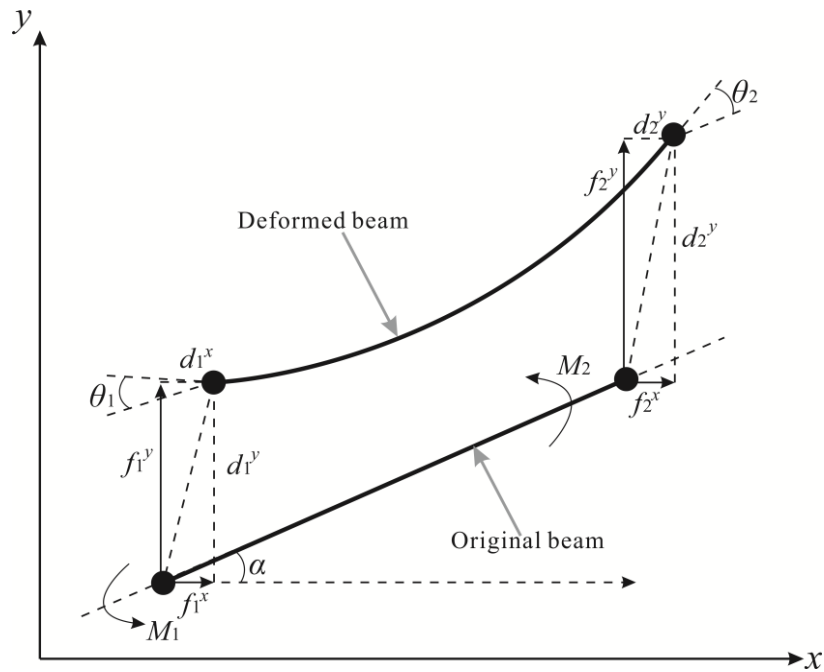

Figure 1. The force analysis of a beam; the figure is redrawn based on Liu et al. (2014). ($f_1^x$, $f_1^y$, $M_1$) and ($f_2^x$, $f_2^y$, $M_2$) are the forces and moments applied to the two end nodes of a beam; ($d_1^x, d_1^y, \theta_1$) and ($d_2^x, d_2^y, \theta_2$) are the displaced distances and rotations of the two end nodes of a beam.

The deformations of the beams are modeled as energy, and their optimal shapes and positions are determined using the principle of minimum energy. Finite Element Modelling (FEM) is used to solve the minimum energy problem. The resulting building locations after displacement are then obtained and the displacement of each building ($d$) is represented as a matrix function, as shown in Eq. (2). For more details about the mathematical derivation process see Bader (2001) and Liu et al. (2014).

$$d = K^{-1}f \quad (2)$$

where $K$ is the element stiffness matrix of each beam; $f$ is the elemental force of each beam, and the detailed representations for $d$, $K$, and $f$ are shown in Eq. (3).

$$\begin{bmatrix} d_{x1} \\ d_{y1} \\ \theta_1 \\ d_{x2} \\ d_{y2} \\ \theta_2 \end{bmatrix} = \frac{l}{E} \begin{bmatrix} Ac^2 + \frac{12I}{l^2}s^2 & (A-\frac{12I}{l^2})cs & -\frac{6I}{l}s & -Ac^2 - \frac{12I}{l^2}s^2 & (A-\frac{12I}{l^2})cs & -\frac{6I}{l}s \\ (A-\frac{12I}{l^2})cs & As^2 + \frac{12I}{l^2}c^2 & \frac{6I}{l}c & -(A-\frac{12I}{l^2})cs & -As^2 - \frac{12I}{l^2}c^2 & \frac{6I}{l}c \\ -\frac{6I}{l}s & \frac{6I}{l}c & 4I & \frac{6I}{l}s & -\frac{6I}{l}c & 2I \\ -Ac^2 - \frac{12I}{l^2}s^2 & -(A-\frac{12I}{l^2})cs & \frac{6I}{l}s & Ac^2 + \frac{12I}{l^2}s^2 & (A-\frac{12I}{l^2})cs & \frac{6I}{l}s \\ -Ac^2 - \frac{12I}{l^2}s^2 & -As^2 + \frac{12I}{l^2}c^2 & -\frac{6I}{l}c & (A-\frac{12I}{l^2})cs & As^2 + \frac{12I}{l^2}c^2 & -\frac{6I}{l}c \\ \frac{-6I}{l}s & \frac{6I}{l}c & 2I & -\frac{6I}{l}s & -\frac{6I}{l}c & 4I \end{bmatrix}^{-1} \begin{bmatrix} f_1^x \\ f_2^y \\ M_1 \\ f_2^x \\ f_2^y \\ M_2 \end{bmatrix} \quad (3)$$

where $A$, $I$, and $E$ are the parameters that describe the material properties of the beams and can be set by users, $A$ controls the stretching and compression property of the beam under axial load, $I$ controls the bending behavior, and $E$ is used to adjust the elasticity of the beams; $l$ is the length of a beam; $\alpha$ is the angle spanned by the beam axis and the x-axis, $s = \sin\alpha$ and $c = \cos\alpha$. We will explain the choices of the parameter values $A$, $E$, and $I$ in *Section* 3.4.

## 3. Methodology

Our approach comprises four steps:

Step 1: Generate initial circles proportional to statistical values;

Step 2: Build a proximity graph for circles based on the relationships between their represented regions;

Step 3: Compute forces for the proximity graph according to the quality requirements in *Section* 2.1;

Step 4: Displace the circles using the elastic beam algorithm and assess whether an acceptable result is achieved; if so, stop; otherwise, return to Step 2.

We provide further details for each of these steps in the following sections.

### 3.1 *Step* 1: *Initial circle generation*

The initial circles are generated to represent their corresponding regions with their locations in the centers of the regions and their areas being proportional to the data values (Dorling, 1996). The radius ($R_i$) of a circle that represents $reg_i$ with value $V_i$ is defined as Eq. (4).

$$R_i = \sqrt{V_i / V_{\min}} * R_{\min} \qquad (4)$$

where $V_{\min}$ is the minimum value of all regions, $R_{\min}$ is the minimum radius of the circles which can be set based on the minimum resolution of human eyes, e.g., 1 mm in graphics (He and Song, 2015).

**3.2 Step 2: *Proximity graph construction for the circles***

The proximity graph is the basic structure for the elastic beam algorithm, and it should be constructed according to quality requirements (Liu et al., 2014). Suppose the regions are $regS = \{reg_1, reg_2, ..., reg_m\}$, the circles representing the regions are $CS = \{C_1, C_2, ..., C_m\}$. The proximity graph ($G$) for $CS$ can be represented as $G = (E, V)$, $V = \{v_1, v_2, ..., v_m\}$ is a node set, $v_m$ represents the circle $C_m$, and $E = \{e(v_i, v_j), v_i, v_j \in V\}$ is the edge set. $G$ is first built by considering the topology relation maintenance, and two refinements are then implemented by considering overlap avoidance and relative relation maintenance.

(1) Proximity graph construction by considering topology relation maintenance

Topology relation maintenance rules that the adjacent relations between regions need to be maintained as much as possible. Thus, the initial proximity graph ($G$) is constructed for the circles by considering the adjacent relations between their represented regions as follows: <u>For each pair of adjacent regions ($reg_i$ and $reg_j$), an edge $e(v_i, v_j)$ connecting $C_i$ and $C_j$ is inserted into $G$.</u> The initial $G$ for the regions in Figure 2(a) can be built as Figure 2(b), where the length ($L$) of $e(v_i, v_j)$ is defined as the distance between its two connected circles ($C_i$ and $C_j$), given by Eq. (5).

$$L = Dis(C_i, C_j) - (R_i + R_j) \qquad (5)$$

where $Dis(C_i, C_j)$ is the distance between the centers of $C_i$ and $C_j$, $R_i$ and $R_j$ are the radii of $C_i$ and $C_j$.

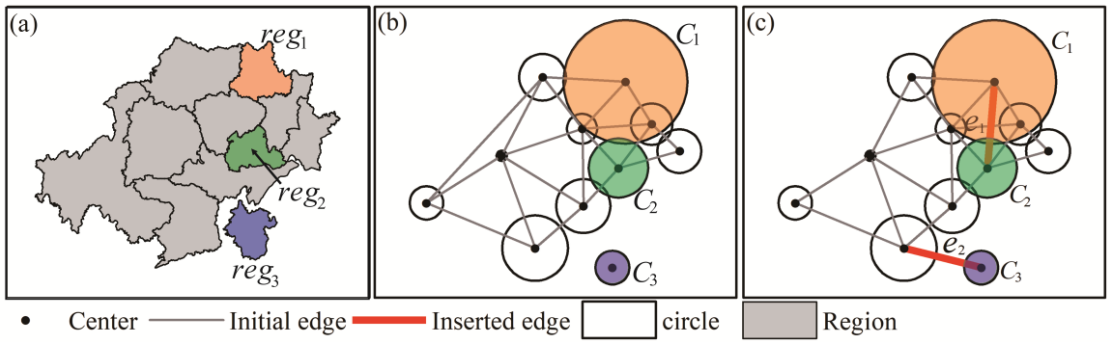

Figure 2. Proximity graph construction. (a) The regions; (b) The built initial proximity graph by considering topology relation maintenance; (c) Insert edges by considering overlap avoidance and relative relation maintenance.

(2) Insert edges by considering overlap avoidance

Overlaps between circles must be avoided (Dorling, 1996). But the circles in $CS$ that overlap each other may not all be initially connected by an edge in $G$. If two circles ($C_i$ and $C_j$) in $CS$ overlap and are not connected by an edge in $G$, the edge $e(v_i, v_j)$ is then inserted into $G$. For example, the edge $e_1$ in Figure 2(c) is inserted because the circles $C_1$ and $C_2$ overlap with each other.

(3) Insert edges by considering relative relation maintenance

The relative relations between circles need to be maintained as much as possible (Inoue, 2011). By using the elastic beam algorithm to calculate forces for each circle, we can achieve global optimization (Bader, 2001; Liu et al., 2014). When connecting all circles as a group, their displacements become interdependent, which helps to preserve the relative relationships between regions. However, the circles may not form a connected component initially, since $G$ is built based on adjacent relations between regions, as in Figure 2(b). Suppose the connected components are $ConCS = \{ConC_1, ConC_2, ..., ConC_n\}$ after the initial $G$ is built, then an edge connecting two circles separately in $CC_m$ and $CC_n$ with the minimum $L$ is inserted into $G$ iteratively until all circles in $CS$ are connected by $G$ as a group. For example, the edge $e_2$ in Figure 2(c) connecting $C_3$ is inserted into $G$ because $C_3$ is disconnected from any other circles.

(4) Edge classification

The edges in $G$ can be classified into two types for later force computation by considering the adjacent relations between regions.

- **Type 1**: Edges that connect two circles whose represented regions are adjacent to each other, e.g., the edges in Figure 2(b);
- **Type 2**: Edges that connect two circles whose represented regions are not adjacent to each other, e.g., $e_1$, $e_2$ in Figure 2(c).

The proximity graph construction for circles is implemented using **Algorithm 1**.

**Algorithm 1. Proximity graph construction**

Input: the regions $regS = \{reg_1, reg_2, ..., reg_m\}$, their represented circles $CS = \{C_1, C_2, ..., C_m\}$

Output: the proximity graph $G = (E, V)$

**Initialize** the output proximity graph as $G = (E, V)$, where $V = \{v_1, v_2, ..., v_m\}$ is the node set, $v_m$

is the center of $C_m$; $E$ is the edge set and $E=Null$

**For each** node pair $(v_i, v_j), v_i \in V, v_j \in V$ **Do**

    **If** $C_i$ and $C_j$ overlap each other or $reg_i$ and $reg_j$ have an adjacent relation **Then**

        add $e(v_i, v_j)$ to $E$

Determine the connected components and denote them as $ConCS = \{ConC_1, ConC_2, ..., ConC_n\}$

**While** (the number of components in *ConCS* is greater than 1) **Do**

    Add the edge that connects two connected components in *ConCS* using the minimum $L$

    Update $G$ and *ConCS*

**Return** $G = (E, V)$

### 3.3 *Step* 3: *Force computation*

Two kinds of forces need to be computed for the proximity graph ($G$), the two circles connected by an edge in $G$ that have an overlap are repulsed by each other to avoid the overlap (**repulsive force**), and the two separated circles that are connected by a Type 1 edge in $G$ are attracted to each other to maintain the topology relation between their corresponding regions (**attractive force**). In this regard, a node in $G$ may be subject to several forces which should be combined into a resultant force. Let $C_i$ and $C_j$ be two circles, with radii $R_i$ and $R_j$, an edge connecting $C_i$ and $C_j$ of length $L$, and centers $O_i$ and $O_j$.

(1) Repulsive force

If $C_i$ and $C_j$ have an overlap, they will accept repulsive forces ($ref_i$ and $ref_j$) to eliminate the overlap, and are defined as Eqs. (6) and (7), as shown in Figure 3(a).

$$ref_i = \frac{\overrightarrow{O_jO_i}}{\left|\overrightarrow{O_iO_j}\right|} * \frac{R_j}{R_i+R_j} * |L| \tag{6}$$

$$ref_j = \frac{\overrightarrow{O_iO_j}}{\left|\overrightarrow{O_iO_j}\right|} * \frac{R_i}{R_i+R_j} * |L| \tag{7}$$

(2) Attractive force

If $C_i$ and $C_j$ are connected by a Type 1 edge in $G$, they will accept attractive force ($atf_i$ and $atf_j$) to maintain the topology relation between their corresponding two regions and are defined as Eqs. (8) and (9), as shown in Figure 3(b).

$$atf_i = \frac{\overrightarrow{O_iO_j}}{\left|\overrightarrow{O_iO_j}\right|} * \frac{R_j}{R_i+R_j} * L \tag{8}$$

$$atf_j = \frac{\overrightarrow{O_jO_i}}{\left|\overrightarrow{O_iO_j}\right|} * \frac{R_i}{R_i + R_j} * L \qquad (9)$$

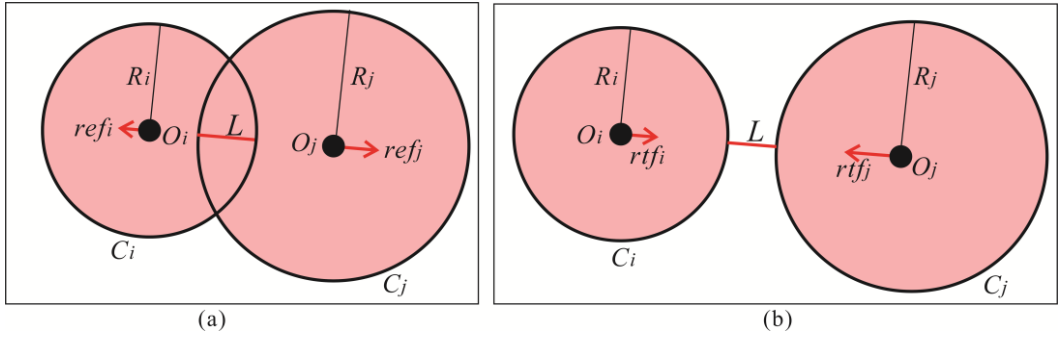

Figure 3. Force computation. (a) Repulsive forces; (b) Attractive forces.

(3) Force combination

A circle may be affected by more than one individual external force (i.e., repulsion or attraction). If these forces are combined into the resultant force by using a simple vector-based summing, inappropriate displacement may occur due to their accumulative effect in a similar direction. Take the circle $C_j$ in Figure 4(a) as an example, it has two forces $f_{ij}$ and $f_{kj}$. A coordinate system can be established by taking the direction of $f_{ij}$ as the main direction, and $f_{kj}$ can then be decomposed into two sub-forces on the x-axis and y-axis as $f^x_{kj}$ and $f^y_{kj}$. If we use the simple vector-based summing method, the two forces $f^x_{kj}$ and $f_{ij}$ on the x-axis are added together. However, only the larger force ($f_{ij}$) is required. To address this, we combine these forces by considering the local maximum component forces in four main directions, such as force $f_{ij}$ on the x-axis and $f^y_{kj}$ on the y-axis in Figure 4(b), and a resultant force can then be obtained. For more details about the force combination please refer to Liu et al. (2014).

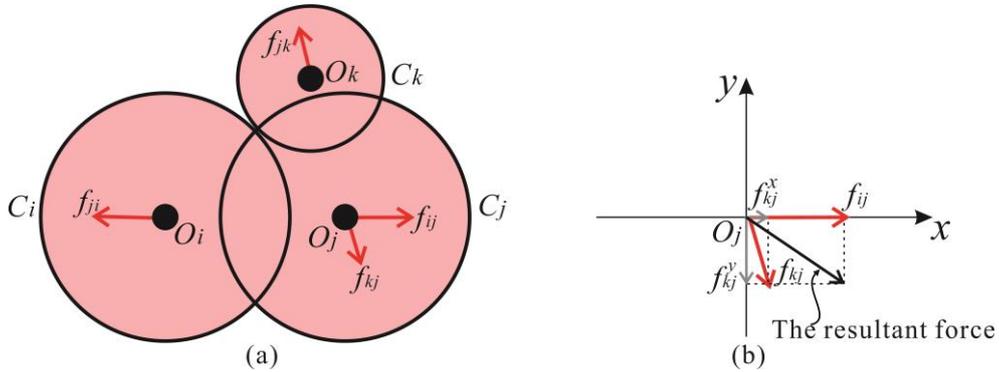

Figure 4. Example for force combination. (a) Forces for each circle; (b) The force combination for circle $C_j$ (Liu et al., 2014).

### 3.4 *Step 4*: *Using the elastic beam algorithm*

The displacement of each node in the proximity graph ($G$) is computed using the elastic beam algorithm. Running the algorithm only once cannot resolve all conflicts that violate the quality requirements or prevent new conflicts from arising, particularly in complex situations (Liu et al., 2014). Therefore, an iterative strategy is applied until all circles are in suitable locations. Furthermore, attractive forces are assigned to ensure that all topology relations between regions are maintained. However, it is not always possible to maintain all topology relations in practice. To address this problem, we relax the requirements for maintaining topology relations after a set maximum step ($Ts$) is reached.

(1) Parameter setting for the elastic beam algorithm

The parameters including $E$, $A$, and $I$ need to be set for the elastic beam algorithm, as described in *Section* 2.3. According to the analysis by Liu et al. (2014), we set $A = L = 1$ to give the beam the same bending and compression properties and reduce the complexity of influencing factors. $E_{k+1}$ is set according to the maximum forces as $\max(f_k)$ for all nodes in the previous iteration of the elastic beam algorithm as follows:

$$E_{k+1} = \frac{\max(d_0)}{\max(f_k)} * E_0 \tag{10}$$

where $\max(d_0)$ means the maximum displaced distance for all nodes in the first iteration of the elastic beam algorithm, and $E_0=10$ according to Liu et al. (2014).

(2) The iterative process

The elastic beam algorithm is applied iteratively to the proximity graph ($G$) to compute the displacement for each node in $G$. The iterative process stops if $\max(f_k) \leq \varepsilon$, where $\varepsilon$ is a small constant, i.e., $\varepsilon=0.001$. Because not all topology relations can be maintained, some attractive forces may be redundant after a set maximum step ($Ts$) is reached. At this time, the circles have been attracted to be as close to each other as possible with the applied attractive forces. If $Ts$ is reached and a satisfactory result cannot be obtained, the attractive forces are not considered in later iterative steps. The iterative process is implemented using **Algorithm 2**.

---

**Algorithm 2. The iterative process for circle displacement using the elastic beam algorithm**

Input: the regions $regS = \{reg_1, reg_2, ..., reg_m\}$, their represented circles $CS = \{C_1, C_2, ..., C_m\}$, and the maximum step with considering the attractive forces is $Ts$

Output: $CS = \{C_1, C_2, ..., C_m\}$ after displacement

Initialize the iterative step, $t \leftarrow 1$

**Do**

  Construction of the proximity graph ($G$)

  **If** $t \leq T_s$ **Then** force computation with attractive forces

  **Else Then** force computation without attractive forces

  Compute the displacement vector for each node in $G$ with the elastic beam algorithm, with the maximum force as $\max(f_k)$;

  Update the coordinates of each $C_m \in CS$ based on the displacement vector

  $t \leftarrow t+1$

**While** ( $\max(f_k) \leq \varepsilon$ )

**Return**: $CS = \{C_1, C_2, ..., C_m\}$

## 4. Experiments

### 4.1 *Implement details and results*

(1) Datasets

  We use two benchmark datasets for cartogram production that have been previously used in the literature (Inoue et al., 2006; Protovis, 2010; Sun et al., 2020; Wei et al., 2022). (1) Dataset A, which includes the population of 48 states of the United States of America in 2015, excluding Alaska and Hawaii (data source: https://hub.arcgis.com/); (2) Dataset B, which includes the population of each country in North and South America in 2021 (data source: https://www.unfpa.org/data/world-population-dashboard).

(2) Parameter settings

  Parameters are set as follows. $R_{\min}$ in Eq. (4) is set as 1 mm according to the minimum resolution of human eyes, and 1mm is a graphic distance (He and Song, 2015). The graphic distance refers to the apparent distance between two points on a map, rather than the actual physical distance on the ground. $T_s$ for the iterative process of using the elastic beam algorithm is set as Eq. (11) by considering the number of the circles ($CN$) according to Liu et al. (2014).

$$T_s = \begin{cases} 20 & (CN \leq 10) \\ 2*CN & (10 < CN < 50) \\ 100 & (CN \geq 50) \end{cases} \quad (11)$$

(3) Experimental environments

  We use C# code to implement our proposed approach based on ArcEngine 10.2, and the experiments are carried out on a personal computer with an Intel® Core™ 1.60 GHz i5-8265U CPU and 8G RAM.

  Two circular cartograms for the two datasets are successfully generated by using the proposed approach, as shown in Figures 5 and 6.

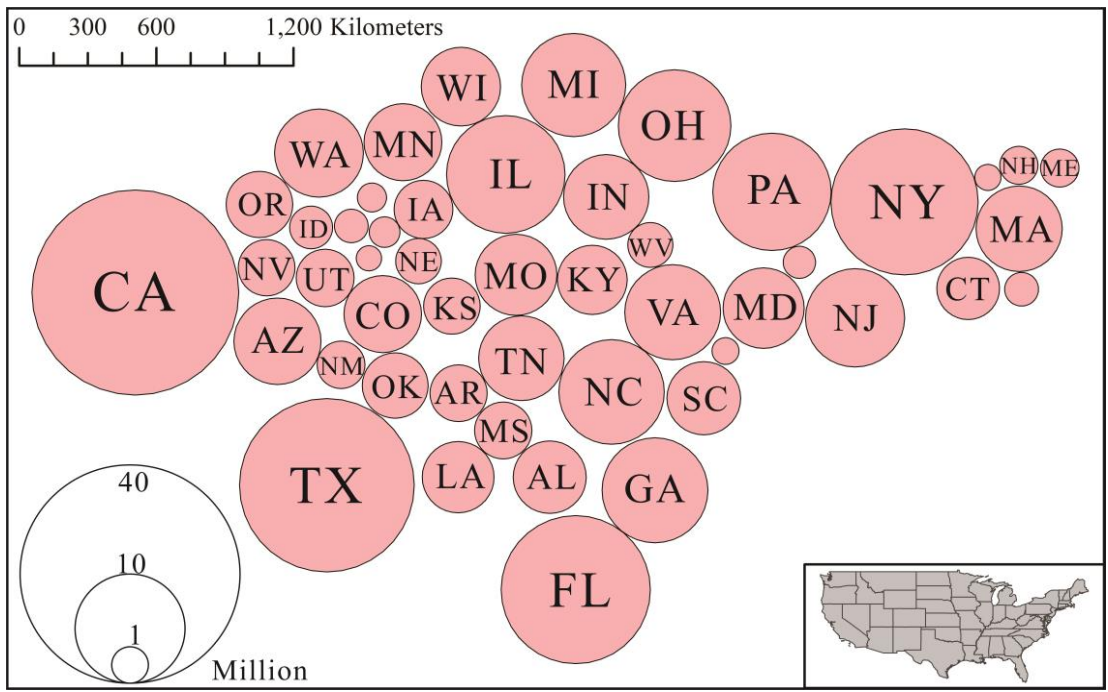

Figure 5. The circular cartogram by using the proposed approach for Dataset A.

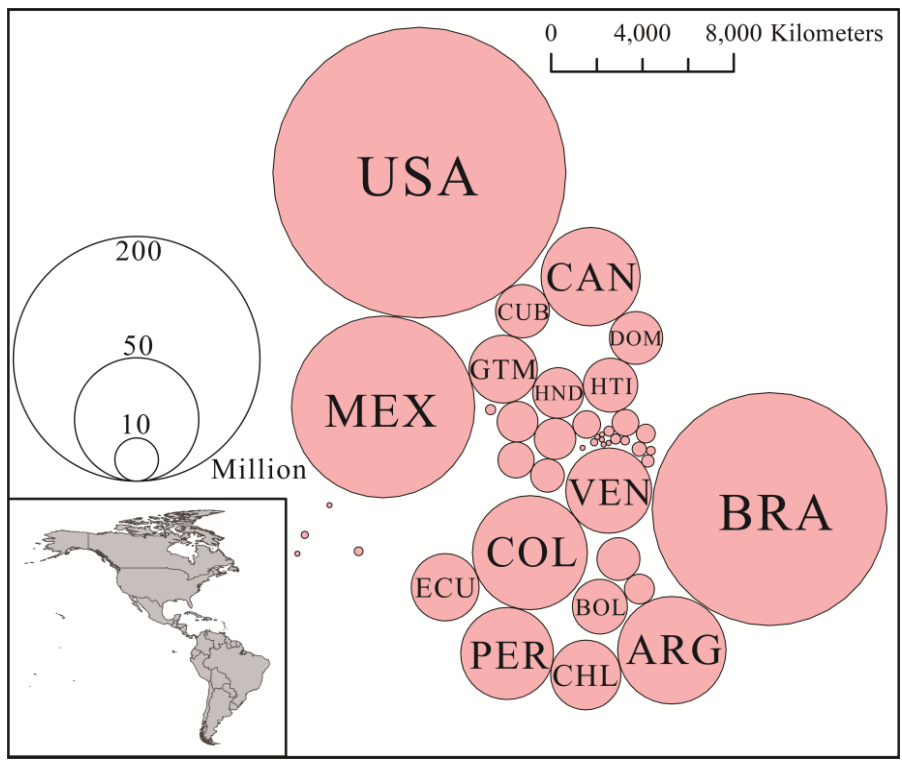

Figure 6. The circular cartogram by using the proposed approach for Dataset B.

### 4.2 *Evaluation and comparison*

4.2.1 *Evaluation*

(1) Effectiveness evaluation

The effectiveness of the proposed approach is evaluated by the following measures according to the quality requirements in *Section* 2.1: (1) The number of overlaps (*NumO*); (2) The ratio of maintained adjacent relations (*RT*); (3) The root mean square (*RMS*) of the difference in direction angles for the links between the centers of neighboring circles (i.e., circles connected by an edge in the constructed proximity graph) (4) The total geographic displacement distance (*TDD*).

The statistical results are shown in Table 1. From Figures 5 and 6 and Table 1, we have the following observations: (1) The regions are represented by circles with their areas in proportion to the populations in the two circular cartograms; (2) No overlaps are generated, indicating that the proposed approach successfully avoids overlaps; (3) The *RT* value for Dataset A is 68.81%, demonstrating that most of the topology relations between regions are maintained; while the *RT* value for Dataset B is 38.46%, it happens due to a large variation in the population of the countries in North and South America; (4) The *RMS* values for Dataset A and Dataset B are 34.61 and 48.35, indicating that the changes in the relative relations between circles are controlled within a low level. Based on these observations, we can conclude that the proposed approach can generate high-quality circular cartograms for various types of data.

Table 1. Statistical results on the effectiveness.

|  | The proposed approach | | Dorling's (1996) approach | |
|---|---|---|---|---|
|  | Dataset A | Dataset B | Dataset A | Dataset B |
| *NumO* | 0 | 0 | 0 | 0 |
| *RT* | 75/109 = 68.81% | 15/39 = 38.46% | 64/109=58.71% | 14/39=35.90% |
| *RMS* | 34.61 | 48.35 | 39.90 | 51.29 |
| *TDD* ($10^6$ m) | 351.81 | 945.85 | 330.16 | 927.18 |

(2) Efficiency evaluation

The efficiency of the proposed approach is evaluated by time consumption (*t*), as shown in Table 2. It will cost 36.74s to create a circular cartogram for Dataset A, whereas, for Dataset B, it is 21.39s. It is important to note that the number of circles (*CN*), the number of edges (*EN*), and the iterative steps (*IS*) of using the elastic beam algorithm are the key factors that affect the efficiency. By comparing Dataset A and Dataset B, we can see that the larger the *CN*, *EN*, and *IS*, the larger the time consumption.

Table 2. Statistical results on the efficiency.

|  | *CN* | *EN* | *IS* | *t(*s) |
|---|---|---|---|---|
| Dataset A | 49 | 114 | 98+34 | 36.74 |
| Dataset B | 43 | 69 | 86+31 | 21.39 |

*4.2.2 Comparison*

To validate the feasibility of the proposed approach, we compare it with Dorling's (1996) approach. A similar Dorling approach is implemented in which topology relations are also maintained as much as possible. The results are shown in Table 1, and we have the following observations. No overlaps are generated using the two approaches, which indicates that both approaches can avoid overlaps. Comparing the proposed approach to Dorling's approach, the *RT* is 10.10% higher for Dataset A and 2.56% higher for Dataset B, the *RMS* is 5.29 less for Dataset A and 2.94 less for Dataset B, and the *TDD* increases $21.65 \times 10^6$m(6.5%) for Dataset A and $18.67 \times 10^6$m (2.0%) for Dataset B. These indicate that the proposed approach can better maintain the topology relations and relative relations between regions, but our approach also results in a larger total displacement distance in comparison to Dorling's approach.

**4.3 *Efficiency scaling analysis***

The efficiency scaling is analyzed by applying the proposed approach to generate circular cartograms for 8 datasets of different areas with the same parameter settings. The 8 datasets are subsets of the 3108 counties in the United States of America (USA) excluding Alaska and Hawaii (see Figure 7). The population of each county in 2017 is used as the statistical values (data source: hub.arcgis.com/datasets/esri::usa-counties-generalized/explore?showTable=true).

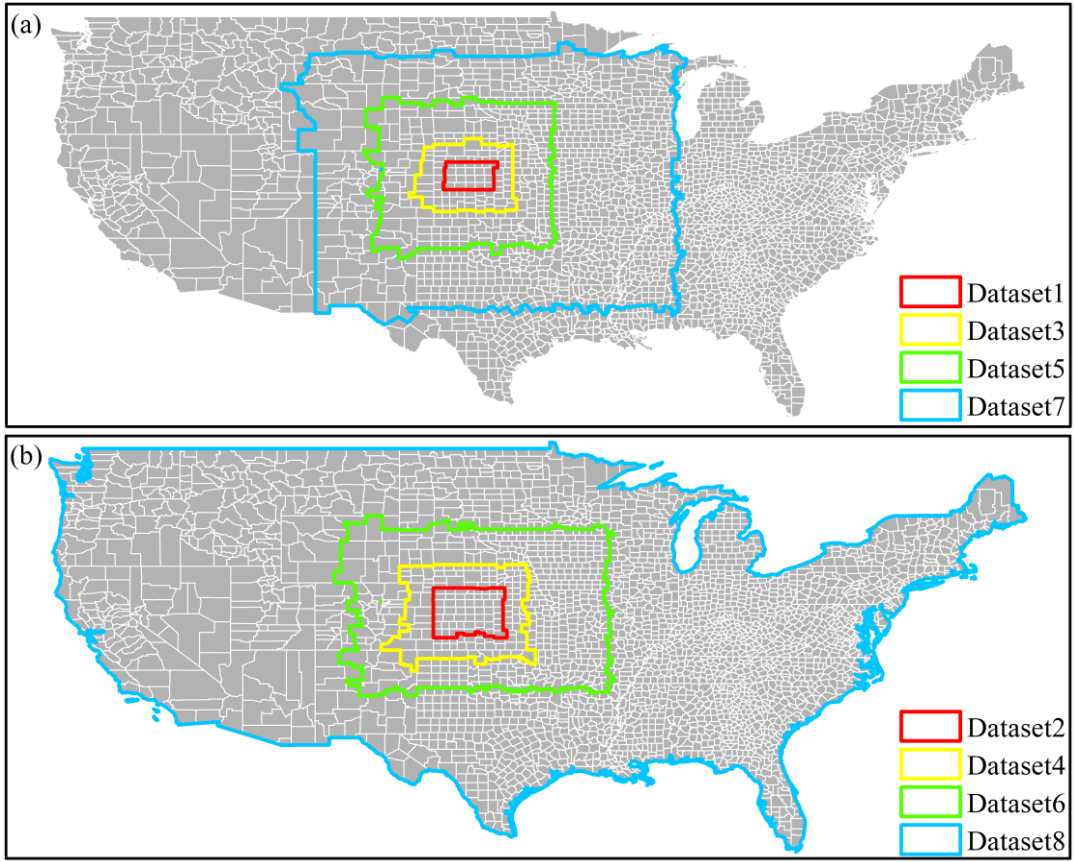

Figure 7. The 8 datasets for efficiency analysis, which are subsets of the 3108 counties of the United States of America (USA) excluding Alaska and Hawaii, the statistical values are the population of each county in 2017.

In Table 3, we present the statistical results for the number of circles (*CN*), the iterative steps (*IS*) of using the elastic beam algorithm, and the time consumption (*t*) on these experimental results. Our observations from the data are as follows: (1) The calculation takes several minutes while $CN \leq 100$. Therefore, the efficiency of the proposed approach may be acceptable if the dataset is not very large. (2) The time consumption (*t*) increases with the increase of *CN* (see Figure 8(a)) due to the proximity graph's many matching and searching operations and the time-consuming matrix operations, as defined in Eq. (3). (3) The small differences in *IS* indicate that it is not the key factor influencing efficiency. The reason is that we set a threshold for *IS*, as discussed in *Sections* 3.4 and 4.1.

Based on the observations and analysis presented above, it is evident that *NC* plays a crucial role in determining efficiency. To enhance efficiency in the future, the following strategies may be implemented: (1) the number of circles used in each application of the elastic beam algorithm could be reduced. For instance, Tang (2013) proposed a method to accelerate circular cartogram production by dividing it into subtasks. Similarly, our approach could benefit from dividing all circles into appropriate sub-groups before its application. (2) To speed up the node or edge matching or the matrix computation in our algorithm, a key-value mechanism could be adopted to represent the proximity graph. This modification would significantly reduce the node or edge matching time.

Table 3. Statistical results on the efficiency scaling.

|  | Dataset1 | Dataset2 | Dataset3 | Dataset4 | Dataset5 | Dataset6 | Dataset7 | Dataset8 |
|---|---|---|---|---|---|---|---|---|
| *CN* | 21 | 54 | 126 | 177 | 291 | 593 | 1363 | 3108 |
| *IS* | 42+12 | 100+27 | 100+63 | 100+48 | 100+92 | 100+80 | —— | —— |
| *t*(s) ↓ | 1.93 | 20.62 | 633.69 | 1524.70 | 9705.15 | 71328.15 | —— | —— |

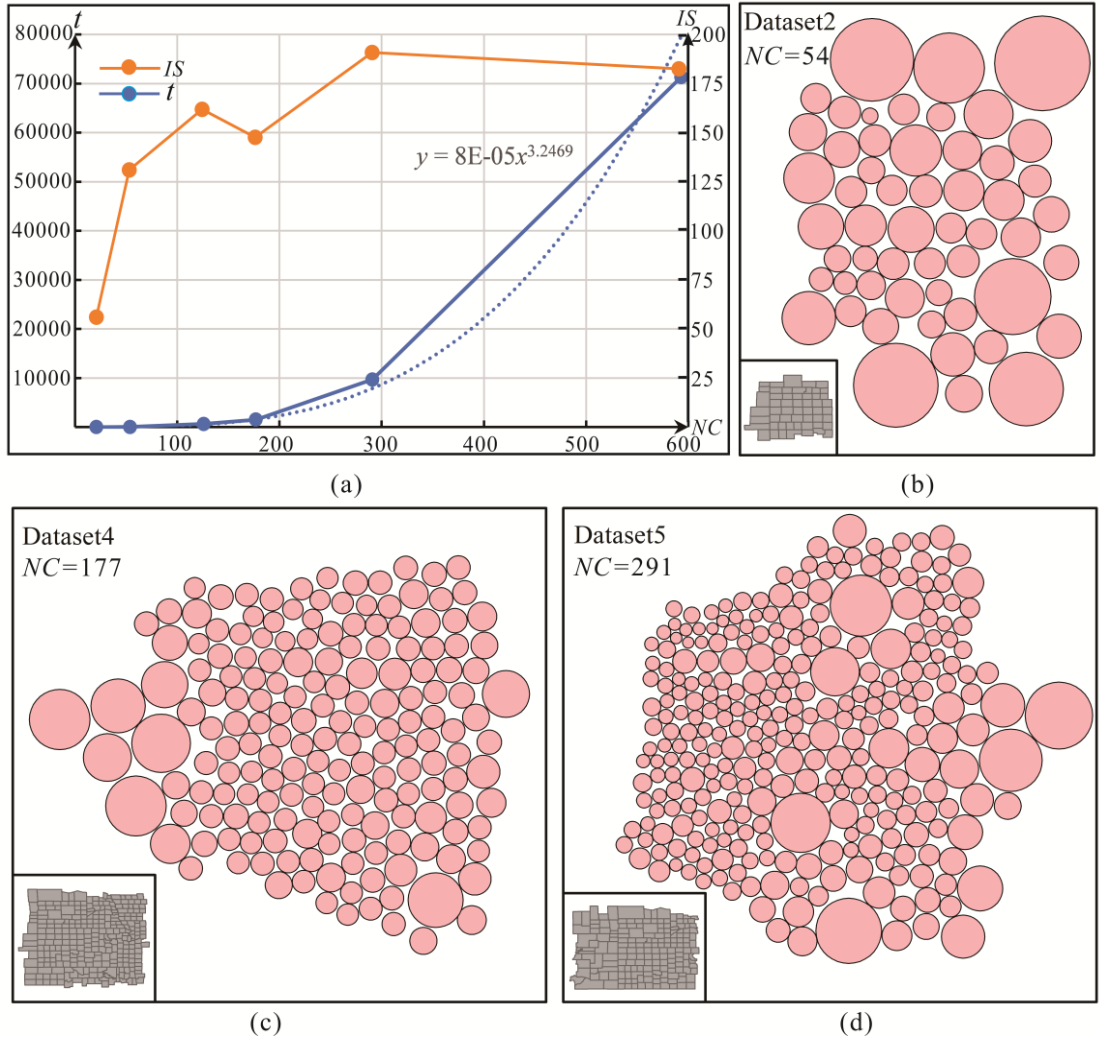

Figure 8. Efficiency scaling analysis. (a) Statistical results on the efficiency scaling. (b), (c), (d) Circular cartograms for Dataset2, Dataset4, and Dataset5.

## 5. Extensions

Although we have summarized the quality requirements of circular cartograms in *Section* 2.1, it is important to note that some specific quality requirements may also need to be considered based on user demands. For example, the users may want to generate a minimum spanning tree (MST) like arrangement by reducing the number of connections between the circles in which circles are connected as a group with the least edges, this may make it easier for the users to better focus on patterns or trends of interest within the data (Wei et al., 2022). The proximity graph is the key to modeling the quality requirements in the proposed approach, and specific quality requirements can be satisfied by constructing a corresponding proximity graph. As user demands are diverse, we here analyze two strategies for proximity graph construction : (1) adding restrictions on the long edges, and (2) applying a specific proximity graph.

(1) Restrictions on the long edges

Two circles connected by a long edge may not be visually close to each other (Wei et al., 2018). Additionally, large displacements may also be required due to these long edges. Therefore, some users may add restrictions on the long edges by specifying a threshold ($T_l$) for the edge length ($L$). If any edge in $G$ satisfies $L > T_l$, then the edge is considered long and will not have attractive forces on the two connected circles. Figure 9 displays a circular cartogram with $T_l$ set to 0, which means topology relation maintenance is not specifically considered. The long edges for evaluation in Figure 5 are not deleted, which means $T_l$ is set as a large constant, and all adjacent relations between regions need to be maintained as much as possible. Because $T_l$ is set to avoid large displacements by loosening the requirement of topology relation maintenance. Therefore, the larger the $T_l$, the more adjacent relations between regions are maintained, but it also means a larger total displacement distance and larger deviation in relative orientations. Comparing Figures 5 and 9, we can see that setting $T_l = 0$ yields smaller *TDD* (33.44 × 10$^6$m), *RMS* (19.62), and *RT* (32.11%), which confirms the above analysis. But the two circular cartograms with different $T_l$ values in Figures 5 and 9 can both avoid overlaps, indicating that satisfactory results can also be obtained by adding restrictions on the long edges.

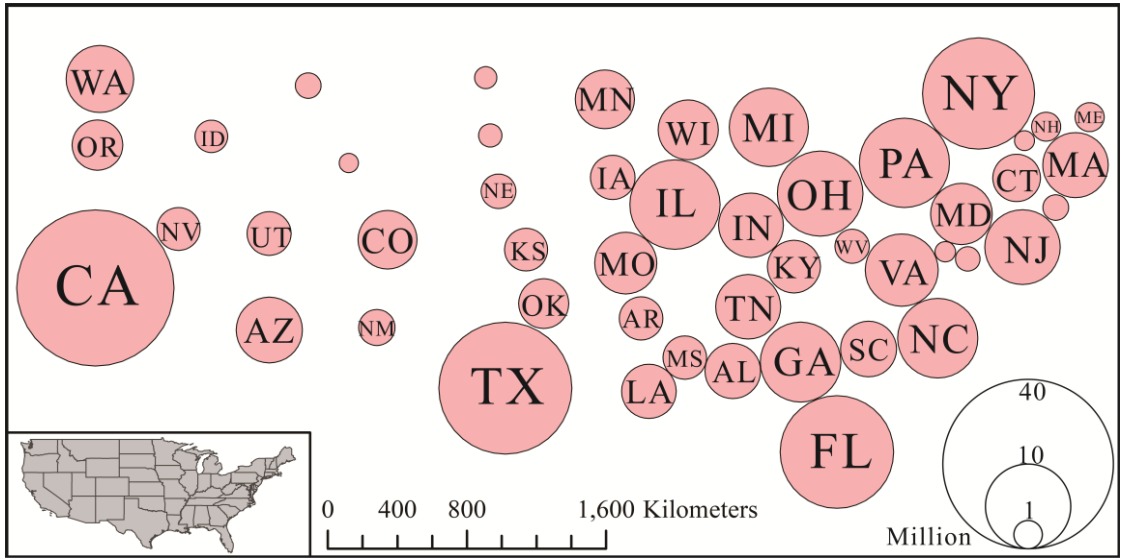

Figure 9. The circular cartograms by using the proposed approach with $T_l = 0$.

(2) Applying a specific proximity graph

In the field of geographic information science, different proximity graphs have been defined to represent various proximity relations, such as the relative proximity graph and minimum spanning tree (MST) (Wei et al., 2018). These proximity graphs can be applied according to specific user demands. For example, in Figure 10, we present a circular cartogram which is generated by using our approach with MST as the proximity graph. In this case, edges are inserted in the MST when two circles overlap, ensuring that the resulting circular cartogram has no overlaps. Since the circles are structured by an MST, their arrangements are also MST-like. The MST can further be changed into any other defined proximity graph ($G$), enabling the generated circular cartogram with a $G$-like arrangement. These results indicate that our proposed approach is applicable to arbitrary proximity graphs.

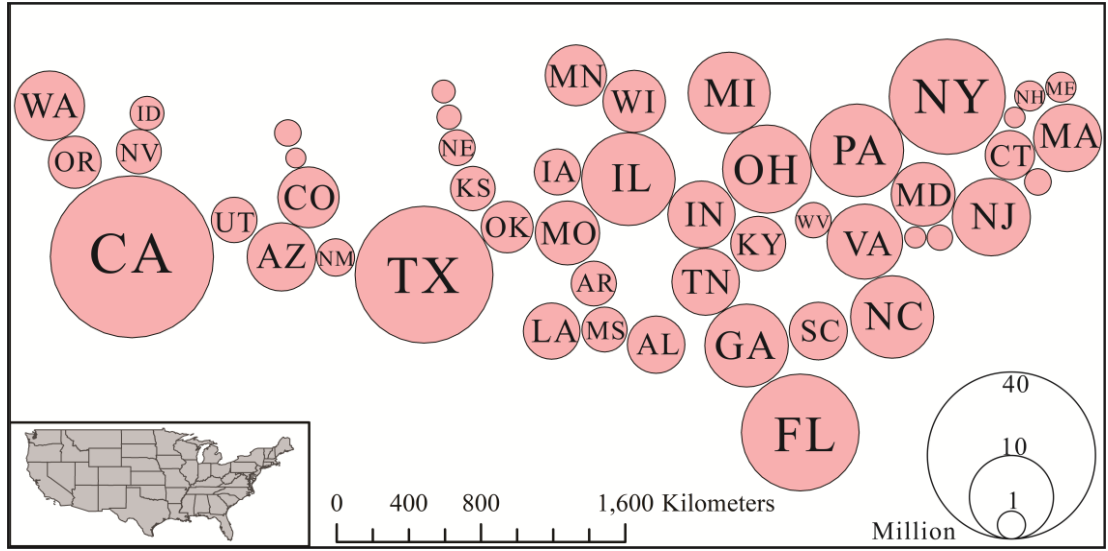

Figure 10. The circular cartogram in which the proximity graph is built based on MST.

## 6. Conclusions

To generate a circular cartogram for quantitative visualization, we employ the elastic beam algorithm, originally used in cartographic generalization, for circle displacements. The displacement mechanism is controlled by enforcing the quality requirements of circular cartograms during proximity graph construction and force computation. Our experimental results indicate that the proposed approach yields higher-quality circular cartograms with better topology and relative relation maintenance than those generated by Dorling's (1996) algorithm. However, the proposed approach also results in larger total displacement distances. In addition, the proposed approach is versatile in accommodating user demands for arbitrary proximity graphs.

Future work will focus on three areas: (1) Conducting a systematic analysis of the utility and usability of circular cartograms, (2) Designing animated circular cartograms that are applicable for time series data, and (3) Improving algorithm efficiency.

**Disclosure statement**

No potential conflict of interest was reported by the author(s).

**Data and code availability statement and data deposition**

The data and code that support the findings of this study are all openly available, website is: https://github.com/TrentonWei/DorlingMap

**Acknowledgements**

This work was supported by the The National Natural Science Foundation of China (grant agreement No 41871378) and The Youth Innovation Promotion Association Foundation of Chinese Academic of Sciences(grant agreement No Y9C0060).

The authors would like to thank Dr. Yuangang Liu, the editors and the anonymous reviewers for their helpful and constructive comments that greatly contributed to improve the manuscript.

**References**

Bader M. (2001) Energy minimization methods for feature displacement in map generalization. Doctorate thesis, Geographic Information System Division, Department of Geography, University of Zurich, Switzerland.

Buchin K, Speckmann B, Verdonschot S. (2012). Evolution strategies for optimizing rectangular cartograms[C]//International Conference on Geographic Information Science. Springer, Berlin, Heidelberg, 29-42.

Bostock M., Ogievetsky V., Heer J. (2011). D3 data-driven documents. IEEE Transactions on Visualization and Computer Graphics 17(12): 2301–2309.

Chrisman, N. R. (1977). Cartogram projections of planar polygon networks. Harvard Laboratory for Computer Graphics and Spatial Analysis, Harvard University.

Chrisman, N. R. (2002). Exploring geographic information systems (2nd ed.). Wiley.

Dorling D F L. Area cartograms: their use and creation[M]//Concepts and techniques in modern geography series. Environmental Publications, University of East Anglia, 1996.

Dougenik, J. A., Chrisman, N. R., Niemeyer, D. R. (1985). An algorithm to construct continuous area cartograms. The Professional Geographer, 37(1), 75–81.

Du, C., Liu, L. (1999). Constructing contiguous area cartogram using ArcView avenue. Proceedings of Geoinformatics '99 Conference, Ann Arbor, MI.

Fei, L. (2002). A method of automated cartographic displacement-on the relationship between streets and buildings (Doctoral dissertation, Univ. Hannover)

Gastner M. T., Newman M. E. J. (2004). Diffusion-based method for producing density-equalizing maps[J]. Proceedings of the National Academy of Sciences, 101(20):

7499-7504.

Gastner M T, Seguy V, More P. (2018). Fast flow-based algorithm for creating density-equalizing map projections[J]. Proceedings of the National Academy of Sciences, 115(10): E2156-E2164.

He Z. & Song Y. (2015). General Map Compilation[M]. Wuhan:Wuhan University Press.

Heilmann R., Keim D., Panse C., Sips M. (2004). Recmap: Rectangular map approximations[C]//IEEE Symposium on Information Visualization. IEEE, 33-40.

Inoue R, Shimizu E. (2006). A new algorithm for continuous area cartogram construction with triangulation of regions and restriction on bearing changes of edges[J]. Cartography and Geographic Information Science, 33(2): 115-125.

Inoue R. (2011). A new construction method for circle cartograms[J]. Cartography and Geographic Information Science, 38(2): 146-152.

Jackel, C. B. (1997). Using ArcView to create contiguous and noncontiguous area cartograms. Cartography and Geographic Information Systems, 24(2), 101–109.

Kronenfeld B. J. (2018). Manual construction of continuous cartograms through mesh transformation. Cartography and Geographic Information Science, 45(1): 76-94.

Liu Y, Guo Q, Sun Y, et al. (2014). A combined approach to cartographic displacement for buildings based on skeleton and improved elastic beam algorithm[J]. PloS one, 9(12): e113953.

Nusrat, S., & Kobourov, S. (2016). The state of the art in cartograms. Computer Graphics Forum, 35(3): 619-642.

Protovis – Dorling Cartograms. http://mbostock. github.io/protovis/ex/cartogram.html, 2010.

Raisz E. (1934). The rectangular statistical cartogram. Geographical Review. 24(3): 292–296.

Reyes Nunez J. J. (2014). The use of cartograms in school cartography. Thematic Cartography for the Society, 327–339.

Robinson A. H. (1955). The 1837 maps of Henry Drury Harness. The Geographical Journal, 121(4): 440–450.

Speckmann B., Kreveld M., Florisson S. (2006). A linear programming approach to rectangular cartograms. In International Symposium on Spatial Data Handling (SDH'06).

Sun, S. (2013a). A fast free-form rubber-sheet algorithm for contiguous area cartograms. International Journal of Geographic Information Science, 27(3): 567–593.

Sun, S. (2013b). An optimized rubber-sheet algorithm for continuous area cartograms. The Professional Geographer, 65(1): 16–30.

Sun S. (2020). Applying forces to generate cartograms: a fast and flexible transformation framework. Cartography and Geographic Information Science, 47(5): 381-399.

Tang W. Parallel construction of large circular cartograms using graphics processing units[J]. International Journal of Geographical Information Science, 2013, 27(11): 2182-2206.

Tobler W. (2004). Thirty-five years of computer cartograms[J]. ANNALS of the Association of American Geographers, 94(1): 58-73.

Van kreveld M., Speckmann B. (2007). On rectangular cartograms. Computational Geometry, 37(3):175–187.

Wei Z., Guo Q., Wang L., F. Yan. (2018). On the spatial distribution of buildings for map generalization. Cartography and Geographic Information Science, 45(6): 539-555.

Wei Z., Xu W., Ding S., Zhang S., Wang Y. (2022). Efficient and stable circular cartograms for time-varying data by using improved elastic beam algorithm and hierarchical optimization[J]. Journal of Visualization, 1-15.

Wolf, E. B. (2005). Creating contiguous cartograms in ArcGIS 9. Proceedings of 2005 ESRI International User Conference, San Diego, CA.